# Coexistence of zigzag antiferromagnetic order and superconductivity in compressed NiPSe$_3$

Hualei Sun[1], Liang Qiu[1], Yifeng Han[3], Enkui Yi[1], Junlong Li[4], Mengwu Huo[1], Chaoxin Huang[1], Hui Liu[1], Manrong Li[3], Weiliang Wang[2], Dao-Xin Yao[1], Benjamin A. Frandsen[5], Bing Shen[1,*], Yusheng Hou[1,#],and Meng Wang[1,†]

[1]*Center for Neutron Science and Technology, Guangdong Provincial Key Laboratory of Magnetoelectric Physics and Devices, School of Physics, Sun Yat-Sen University, Guangzhou, Guangdong 510275, China*

[2] *School of Physics, Guangdong Province Key Laboratory of Display Material and Technology, Sun Yat-Sen University, Guangzhou, Guangdong 510275, China*

[3] *Key Laboratory of Bioinorganic and Synthetic Chemistry of Ministry of Education, School of Chemistry, Sun Yat-Sen University, Guangzhou, Guangdong 510275, China*

[4]*Beijing Synchrotron Radiation Facility, Institute of High Energy Physics, Chinese Academy of Sciences, Beijing 100049, China*

[5]*Department of Physics and Astronomy, Brigham Young University, Provo, Utah 84602, USA*

*# houysh@mail.sysu.edu.cn*

*\* shenbing@mail.sysu.edu.cn*

*† wangmeng5@mail.sysu.edu.cn*

**Abstract**

NiPSe$_3$ is regarded as a bandwidth-controlled Mott insulator, distinct from the widely studied Mott insulating magnetic graphene $M$PSe$_3$ ($M$ = Mn and Fe) family. By employing high-pressure synchrotron X-ray diffraction, we observe two structural transitions as a function of pressure. With the help of first-principles calculations, we discover the antiferromagnetic (AFM) moment directions of NiPSe$_3$ switch from out-of-plane to in-plane and the honeycomb layers slide relative to each other at the first structural transition. The in-plane AFM order persists until the second structural transition, whereupon the two-dimensional (2D) structure assumes a more three-dimensional (3D) character. A bandwidth-controlled Mott insulator-metal transition (IMT) occurs between the two structural transitions at $P_c \approx 8.0$ GPa, concomitant with the emergence of superconductivity with $T_c \approx 4.8$ K. The superconductivity in NiPSe$_3$ emerging in the 2D monoclinic phase coexists with the in-plane AFM order and continues into the 3D trigonal phase. Our electronic structure calculations reveal that the Mott IMT and superconductivity in NiPSe$_3$ are both closely related to the enhanced Se$^{2-}$ 4$p$ and Ni$^{2+}$ 3$d$ electronic hybridizations under pressure. From these results, we construct the temperature-pressure electronic phase diagram of NiPSe$_3$, revealing rich physics and many similarities with copper oxide and iron-based superconductors.

**Introduction**

The $M$PSe$_3$ family, where $M$ is a transition metal, has attracted extensive attention for their unique magnetic properties and potential for spintronic device applications. The long-range AFM order in $M$PSe$_3$ can be maintained even down to the monolayer scale, such that these materials have been described as "magnetic graphene"[1,2,3,4,5]. This is possible because the magnetocrystalline anisotropic energy of magnetism counteracts the tendency of thermal fluctuations to destroy 2D magnetic order[6]. The magnetocrystalline anisotropic energy gap results from distortions of the $M$Se$_6$ octahedra and the hexagonal honeycomb structure[7,8]. The anisotropic magnetism also leads to different types of magnetic interactions and diverse magnetic ordered ground states. For example, MnPSe$_3$ exhibits a Néel type AFM with spins pointing parallel to the van der Waals (vdW) plane[9], while FePSe$_3$ possesses zig-zag type AFM with spins perpendicular to the vdW plane[10].

For $M$PSe$_3$ ($M$ = Mn, Fe, and Ni) under atmospheric pressure, the $M$Se$_6$ octahedra have a nearly regular coordination structure. The transition metals Mn, Fe, and Ni are all in the high spin magnetic states. Pressure can induce distortions of the $M$Se$_6$ octahedra and is expected to tune the crystal field, possibly resulting in spin state transitions of the 3$d$ metals. MnPSe$_3$ and FePSe$_3$ both have partially filled $e_g$ and $t_{2g}$ states. The pressure-induced spin state transitions, called spin crossover transitions, are indeed found coincident with a dramatic decrease of the ionic radius of Mn$^{2+}$ and Fe$^{2+}$ ions. Concomitantly, the $d$-$d$ overlap of the two $t_{2g}$ orbitals between the nearest two Mn$^{2+}$ or Fe$^{2+}$ ions causes a Mn or Fe dimer to form[9,10,11]. Thus, the pressure-induced spin state transition and the formation of metallic bonds in MnPSe$_3$ and FePSe$_3$ are accompanied by a structural transition[12,13]. In addition, superconductivity (SC) was found in the non-magnetic state of FePSe$_3$, where the spin of Fe$^{2+}$ is $S$=0. Previous calculations predicted the lattice structures of NiPSe$_3$ to undergo different behaviors during the IMT under pressure, due to the different occupation states of the $e_g$ and $t_{2g}$ orbitals of Ni ions[12,13]. Therefore, it is of highly interesting to explore the properties of NiPSe$_3$ under pressure. However, high pressure studies on NiPSe$_3$ are absent due to the lack of available single crystals.

Here, we report comprehensive high-pressure studies on NiPSe$_3$ single crystals up to 34.0 GPa utilizing synchrotron X-ray diffraction (XRD), electrical resistance measurements, and first-principles calculations. NiPSe$_3$ undergoes two structural transitions at pressures of ~4.0 and ~15.0 GPa, respectively. The first structural transition corresponds to a sliding of the honeycomb layers accompanied by a reorientation of the moments in the AFM zigzag order from out-of-plane to in-plane. The second one is a transition from the monoclinic symmetry high-pressure I (HP-I) phase to the nonmagnetic trigonal symmetry HP-II phase, coincident with a 2D to 3D structural transition. The in-plane AFM order is suppressed gradually in the 2D monoclinic HP-I phase. An IMT occurs at ~8.0 GPa between the two structural transitions, consistent with a bandwidth-controlled Mott transition. Superconductivity appears immediately following the IMT, coexisting with the zigzag AFM order in the HP-I phase and persisting into the nonmagnetic HP-II phase.

**Results**

**Pressure-induced structural transitions in NiPSe$_3$.** The Ni$^{2+}$ ions in NiPSe$_3$ form a 2D honeycomb layered lattice. Each Ni$^{2+}$ ion is octahedrally coordinated by six nearest-neighbor Se atoms. The honeycomb sublattices stack along the $c$ axis. Figure 1a shows the high-pressure powder XRD patterns taken at room temperature up to 25.7 GPa. Two distinct phase transitions can be identified in this pressure range. The first transition occurs at ~4.0 GPa with a new peak appearing at ~13.2°. We call this the HP-I phase. The peak at ~13.7° in the low pressure (LP) phase is suppressed with further increasing pressure. The second transition occurs around 15.0 GPa, as evidenced by a new peak emerging at ~14.4°. The peak at ~13.6° from the HP-I phase disappears quickly under additional pressure. Rietveld refinements performed with TOPAS-Academic[14] are shown for representative pressures in the HP-II, HP-I, and LP phases in Figs. 1b, 1c, and 1d, respectively. Figure 2 displays the refined structures. The detailed structural parameters are listed in Supplementary Table S1.

The structural transition at ~4.0 GPa is an isomorphic structural transition, where the honeycomb layers shift relative to each other in a sliding motion of ~$a$/3 along the $a$-axis, resulting in the $\beta$ angle of the monoclinic unit cell contracting to nearly 90°. Such a large sliding of the honeycomb sublattices is possible due to the weak interlayer vdW interactions. The unit cell remains in the monoclinic space group *C*2/*m*. The second structural transition to a trigonal symmetry (space group *P*-31*m*) at ~15 GPa is non-isomorphic. There is an obvious reduction of the interlayer distance. At 25.7 GPa, the distance between the two nearest P ions is 2.224(3) Å, which is much less than the distance between two vdW-coupled P ions (3.8 Å) and quite close to the phosphate dimer distance (2.21 Å)[15]. This collapse along the *c* axis therefore corresponds to a transition from the quasi-2D layered structure at lower pressure to a genuinely 3D structure at higher pressure. A similar structural transition and 2D-3D crossover in NiPSe$_3$ were recently reported[16].

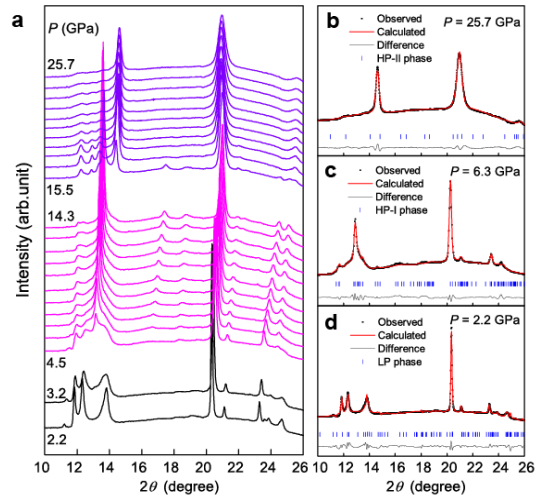

**Fig. 1 a High-pressure XRD patterns from 2.2 to 25.7 GPa.** The X-ray wavelength is λ=0.6199 Å. Two structural transitions occur, one between 3.2 and 4.5 GPa, and the other between 14.3 and 15.5 GPa. **b**-**d** Rietveld refinements at 25.7, 6.3 and 2.2 GPa, corresponding to the HP-II, HP-I, and LP phases.

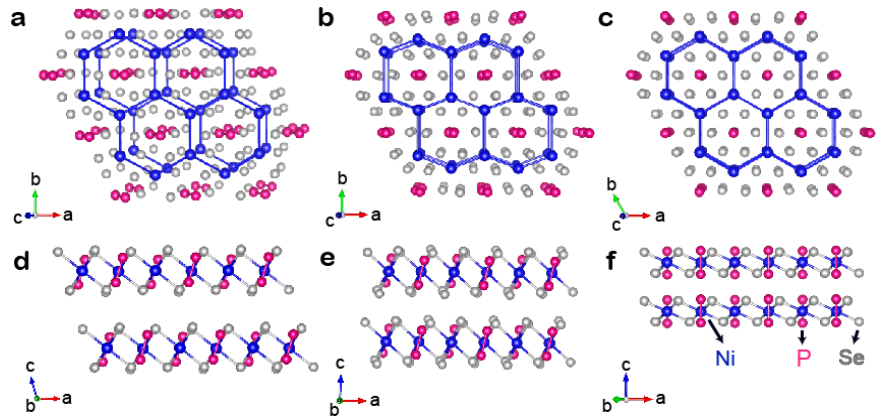

**Fig. 2 Schematics of the structural phases of NiPSe$_3$ under pressure.** (**a, d**) Refined structure in the LP phase at 2.2 GPa, displayed along a viewing axis perpendicular to and parallel to the vdW planes, respectively. The different orientations are drawn to the same scale with respect to the lattice parameters. (**b, e**) Equivalent figures for the HP-I structure at 6.3 GPa. (**c, f**) Equivalent figures for the HP-II structure at 25.7 GPa.

**Pressure-induced IMT and superconductivity.** We now investigate the possibility that changes in electrical transport properties of NiPSe$_3$ accompany the observed structural transitions[17–20]. Figure 3a shows the temperature dependence of the resistance for a single crystal of NiPSe$_3$ measured at high pressures up to 32.0 GPa. We observe a clear IMT under pressure. The resistance as a function of pressure for selected temperatures is presented in Fig. 3b. At all measured temperatures, we observe an abrupt decrease of the resistance at the pressure corresponding to the isomorphic structural transition from the LP phase to the HP-I phase, with NiPSe$_3$ becoming completely metallic at ~ 8.0 GPa, which is between the two structural transitions. The electrical transport measurements have been repeated on several samples (see Supplementary Figs. S2a and S2b).

In conjunction with this IMT, we observed a significant drop in resistance below 4.9 K for an applied pressure of 8.6 GPa. We will show that this corresponds to a SC transition. The transition temperature extracted from the resistance curves initially increases as a function of pressure across the HP-I and HP-II phases, reaches a maximum of 5.9 K around 27.6 GPa, and then remains constant or decreases slightly with higher pressures. This is illustrated in Fig. 3d. To determine whether the drop in resistance represents a SC transition, we measured the resistance in an applied magnetic field in the HP-I phase at 14.0 GPa (Figs. 3e and 3f) and the HP-II phase at 15.6 GPa (Figs. 3g and 3h). The drop in resistance is clearly suppressed to lower temperature with increasing magnetic field, revealing a SC transition. We note that the resistance remains nonzero at low temperature, but this is commonly observed for pressure-induced superconductivity, possibly due to lattice distortions or inhomogeneous pressure[21,22,23,24]. The Ginzburg-Landau formula $\mu_0 H_{c2}(T) = \mu_0 H_{c2}(0)\left[1 - \left(\frac{T}{T_c}\right)^2\right]$ is adopted to fit the upper critical field of the superconductivity. The fitted values of $\mu_0 H_{c2}$ are 0.91 T at 14.0 GPa and 3.95 T at 15.6 GPa, lower than the Pauli limit. We see that both the transition temperature and the upper critical field are enhanced in the HP-II phase (see Supplementary Figs. S2c-S2f).

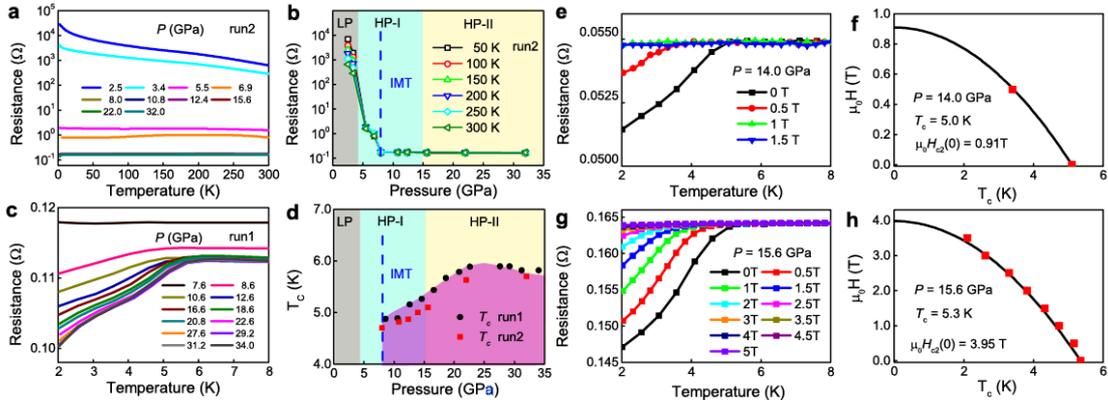

**Fig. 3 Electrical transport measurements under pressure and magnetic field. a** Temperature dependence of the resistance for single-crystal NiPSe$_3$ at pressures between 2.5 to 32.0 GPa. The vertical axis is on a logarithmic scale. The color of each curve indicates the corresponding pressure. **b** Pressure dependence of the resistance at various temperatures up to 300 K. The vertical axis is on a logarithmic scale. The grey, blue, and yellow backgrounds indicate the LP, HP-I, and HP-II phases. The dashed line indicates the IMT at ~ 8.0 GPa. **c** A zoomed-in plot of the resistance as a function of temperature from 2 to 8 K at pressures between 7.6 and 34.0 GPa. **d** SC transition temperature $T_c$ extracted from the measurements of two single crystals at various pressures. **e, g** Magnetic field dependence of $T_c$ at 14.0 and 15.6 GPa. **f, h** Ginzburg-Landau fits to the experimentally determined $T_c$ as a function of magnetic field at 14.0 and 15.6 GPa. The values of the upper critical field $\mu_0 H_{c2}(0)$ and $T_c$ in the absence of an applied magnetic field are labelled.

**DFT calculations of electronic structure and magnetic order.** To obtain the comprehensive electronic structure of NiPSe$_3$ under different pressures, we first investigate the band structure using density functional theory (DFT) calculations. As shown in Fig. 4a, NiPSe$_3$ remains insulating with an indirect gap of 0.316 eV when the pressure is 6.3 GPa. As the pressure increases to 12.3 GPa, there are electronic pockets between Γ and Y but holes between B and E in our DFT calculated band structure (Fig. 4b). Such features indicate that NiPSe$_3$ has become a metal, consistent with the experimentally observed pressure-induced IMT at 8.0 GPa. When the pressure is further increased to 25.7 GPa, many bands cross the Fermi level (Fig. 4c), revealing that NiPSe$_3$ is a good metal. From the projected density of state (PDOS) as shown in Figs. 4d-4f, we can see that the 3$d$ states of Ni$^{2+}$ ions are mainly located below the Fermi level, irrespective of the pressure. It is worth noting that the PDOS from Ni$^{2+}$ and Se$^{2-}$ ions significantly increase near the Fermi level as the pressure increases. This is understandable because the pressure shortens the bond lengths between Ni$^{2+}$ and Se$^{2-}$ ions and thus enhances the $p$-$d$ hybridization. Therefore, the IMT and superconductivity in NiPSe$_3$ are closely related to the enhanced $p$-$d$ hybridization between Ni$^{2+}$ and Se$^{2-}$ ions under pressure. Previous calculations also show that the partially filled $e_g$ orbital of Ni$^{2+}$ is at the Fermi level and contributes to the metallic electrons[25].

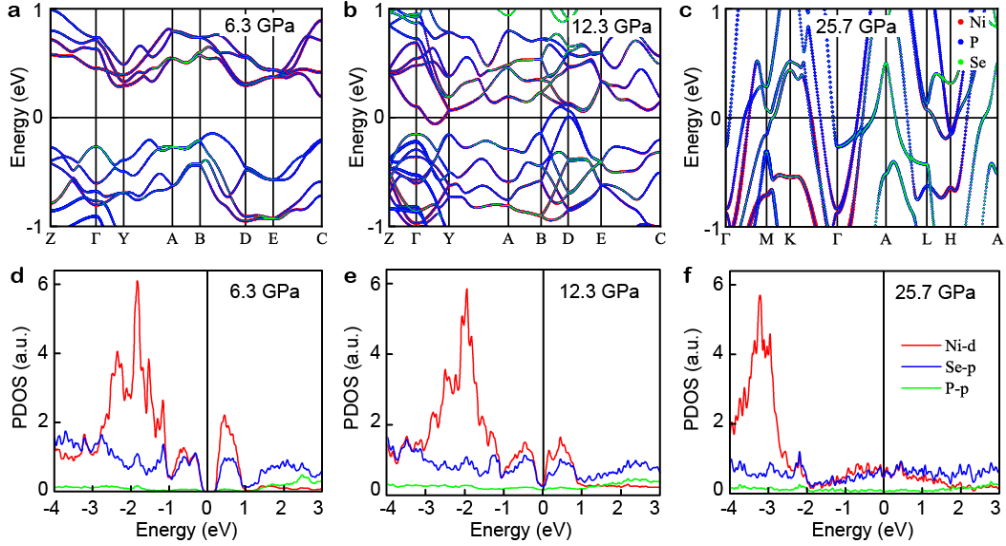

**Fig. 4 DFT-calculated electronic structure of NiPSe$_3$ under different pressures. a-c** Band structure with spin–orbit coupling (SOC) included at 6.3, 12.3, and 25.7 GPa. **d-f** PDOS of Ni 3$d$, Se 4$p$, and P 3$p$ orbitals at 6.3, 12.3, and 25.7 GPa.

To understand the influence of pressure on the magnetic order and the Néel temperature $T_N$ in NiPSe$_3$, we investigate the magnetic exchange couplings based on the following spin model:

$$H = \sum_{ij} J_{ij} \mathbf{S}_i \cdot \mathbf{S}_j + A \sum_i \left( S_i^z \right)^2 \quad (1).$$

Considering the layered honeycomb structure of NiPSe$_3$, the five nearest neighbor (NN) Heisenberg exchange couplings (parameterized by $J_{ij}$) are included. In Eq. (1), $A$ is the single-ion magnetic anisotropy parameter. The values of $T_N$ under different pressures could be identified from both resistance and calculations (see Supplementary Figs. S3 and S4, and Table S2). The calculated $T_N$ at atmospheric pressure is 189.9 K, close to 206 K determined by neutron scattering[26]. As the pressure increases, $T_N$ obtained from the resistance curves decreases from 203.1 K at 2.0 GPa to 139.2 K at 7.6 GPa, as shown in Fig. 5. DFT calculations and Monte Carlo simulations indicate that the magnetic structures are different in the LP and HP-I phases, despite the isomorphic nature of the structural phase transition. For the LP phase, the magnetic moment directions are perpendicular to the honeycomb layers and exhibit an

AFM arrangement within the plane and a FM arrangement between planes (designated as zigzag-out, see the inset in Fig. 5). The gap is 0.99 eV. For the HP-I phase, the magnetic moment directions are within the honeycomb layers, displaying an AFM arrangement both within and between layers (designated as zigzag-in, see the inset in Fig. 5). Lastly, our DFT calculations show that the HP-II phase is non-magnetic. Pressure causes a significant decrease of the interlayer distance, resulting in further enhancement of the metallization and the disappearance of the magnetism. This is confirmed by high pressure Raman scattering in NiPS$_3$, which showed a complete suppression of $T_N$ at the second structural transition[27].

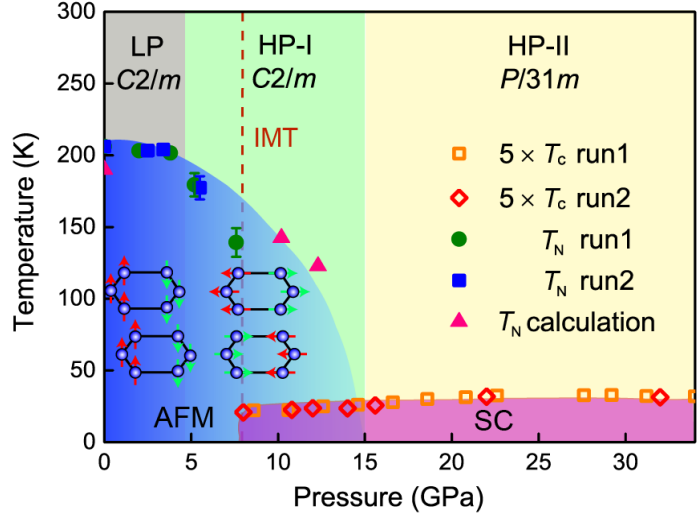

**Fig. 5 Phase diagram of NiPSe$_3$ under pressure.** The insets illustrate the magnetic structures, i.e. zigzag-out for the LP phase and zigzag-in for the HP-I phase. $T_N$ for the AFM order and $T_c$ for the superconductivity are extracted from resistance measurements and DFT calculations.

## Discussion

**Bandwidth controlled insulator-metal transition.** The IMT occurring at ~8.0 GPa in NiPSe$_3$ is separated from the structural transitions. This is distinct from the $d^4$-$d^7$ transition metal complexes, where both the $t_{2g}$ and $e_g$ orbitals of the transition metals are partially filled. For FePSe$_3$ and MnPSe$_3$, the IMT, spin crossover, and structural transition appear simultaneously due to the interorbital $e_g$-$t_{2g}$ hopping. However, Ni$^{2+}$ ($3d^8$) has fully filled $t_{2g}$ orbitals and partially filled $e_g$ orbitals, producing no ionic radius collapse under pressure. Pressure induces the sliding of the vdW layers before the IMT. It was suggested that NiPS$_3$ exhibits negative charge-transfer character with a $d^9 \underline{L}$ configuration where $\underline{L}$ indicates a ligand (chalcogen) hole [28–30]. The IMT can be ascribed to progressive enhancement of the $p$-$d$ hybridization between Ni$^{2+}$ and Se$^{2-}$ ions and increase of the Ni-$d^9$ configuration ($S=1/2$) under pressure[12,13], thus qualifying it as a bandwidth controlled IMT. The 2D monoclinic structure with vdW layers observed in the LP and HP-I phases transforms into a 3D trigonal structure in the HP-II phase, in low-spin $S=0$ Ni ions with itinerant electrons are favored.

**Coexistence of superconductivity and magnetic order.** Our theoretical analysis reveals the LP phase to be a Mott insulator with out-of-plane zigzag AFM order. With increasing pressure, the structure transitions to the HP-I phase characterized by the honeycomb layers sliding relative to each other, and the magnetic moments reorient to point in the plane. Superconductivity emerges from the zigzag AFM parent phase in the HP-I structure. The magnetic order in the SC phase is dominated by the $S=1/2$ AFM order, due to charge transfer from Se under pressure. The magnetism and superconductivity coexist and compete reminiscent of copper oxide and iron-based superconductors. In the HP-II phase, the S=0 state is favored, yielding a nonmagnetic phase.

In summary, our work reveals that pressure induces structural and magnetic transitions in NiPSe$_3$. A bandwidth-controlled insulator-metal transition is observed, accompanied by the emergence of superconductivity. Interestingly, the LP phase and HP-I phase have different zigzag antiferromagnetic orders with magnetic moments perpendicular and parallel to the honeycomb layers. Pressure induces a significant reduction of the interlayer distance, resulting in further enhancement of the metallization and the disappearance of the magnetism. The HP-II phase shows 3D character with a nonmagnetic ground state. The variation of the structural dimension and the cooperation between spin and lattice degrees of freedom make NiPSe$_3$ an interesting compound to explore intertwined superconductivity and magnetism.

**Methods**

**Material synthesis.** NiPSe$_3$ powders were synthesized by heating the stoichiometric Ni, P and Se powders at 600 °C in sealed quartz ampoule for two weeks. High quality single crystals of NiPSe$_3$ were grown with pure NiPSe$_3$ powders mixed with NaCl/AlCl$_3$ using the chemical vapor transport method. The starting materials were sealed in a quartz ampoule and placed in a two-zone furnace for one month. The NiPSe$_3$ single crystals were acquired by dissolving the flux by water. The crystals were naturally cleaved along the (001) surface with shiny gray appearance. The crystals were irregular hexagons with sides of ~1 mm in length and ~10-50 μm in thickness.

**High-pressure XRD measurements.** High-pressure synchrotron radiation XRD patterns of NiPSe$_3$ were collected at 300 K with an X-ray wavelength of 0.6199 Å. A symmetric diamond anvil cell (DAC) with a pair of 300-μm-diameter culets was used. A sample chamber with a diameter of 110 μm was laser-drilled in a pre-indented steel gasket. The NiPSe$_3$ powders were compressed into a pellet with a 60-μm diameter and 20-μm thickness. The pellet was loaded into the middle of the sample chamber and silicone oil was used as a pressure transmitting medium. A ruby sphere was also loaded into the sample chamber and pressure was determined by measuring the shift of its fluorescence wavelength. The data were initially integrated using Dioptas[31] (with a CeO$_2$ calibration) and the subsequent Rietveld refinements were processed using TOPAS-Academic.[14]

**High-pressure electrical property measurements.** Magnetic and electrical measurements were taken on a physical property measurement system (PPMS, Quantum Design). High-pressure electrical transport measurements of NiPSe$_3$ single crystals were carried out using a miniature DAC made from a Be–Cu alloy on a PPMS. Diamond anvils with a 300-μm culet were used, and the corresponding sample chamber (with a diameter of 110-μm) was made in an insulating gasket achieved by cubic boron nitride and epoxy mixture. NaCl powders were employed as the pressure-transmitting medium, providing a quasi-hydrostatic environment. The pressure was also calibrated by measuring the shift of the fluorescence wavelength of the ruby sphere, which was loaded in the sample chamber. The standard four-probe technique was adopted for these measurements.

**First-principles calculations.** DFT calculations are performed using the Vienna *ab initio* Simulation Package (VASP) at the level of the generalized gradient approximation.[32,33] We adopted the projector augmented wave pseudopotentials and a plane-wave cutoff energy of 500 eV.[34] The experimentally measured lattice constants were used in our calculations and the positions of all atoms were fully relaxed until the force on each atom was less than 0.01 eV/Å. We used $U = 2.8$ eV for atmospheric pressure but $U = 0.4$ eV for higher pressures to describe the correlation among $3d$ electrons of $Ni^{2+}$ ions. The values of $T_N$ were obtained through parallel tempering Monte Carlo (MC) simulations.[35,36]

**Acknowledgements**

Work at Sun Yat-Sen University was supported by the National Natural Science Foundation of China (Grant Nos. 12174454, 12104518, U213010013, 11974432, 92165204), Guangdong Basic and Applied

Basic Research Funds (Grant Nos. 2021B1515120015, 2022A1515012643, 2022A1515010035, 2022B1212010008), Guangzhou Basic and Applied Basic Research Funds (Grant Nos. 202201011123, 202201011118, 202201011798), National Key Research and Development Program of China (Grant Nos. 2019YFA0705702, 2022YFA1402802, 2018YFA0306001), Shenzhen International Quantum Academy (Grant No. SIQA202102), the Fundamental Research Funds for the Central Universities, Sun Yat-sen University (Grant No. 22QNTD3004), DFT calculations are performed on Tianhe-II. We appreciate the support of BSRF, IHEP, CAS for high pressure XRD measurements.

# Supplementary Materials for "Coexistence of zigzag antiferromagnetic order and superconductivity in compressed NiPSe$_3$"

Hualei Sun[1], Liang Qiu[1], Yifeng Han[3], Enkui Yi[1], Junlong Li[4], Mengwu Huo[1], Chaoxin Huang[1], Hui Liu[1], Manrong Li[3], Weiliang Wang[2], Dao-Xin Yao[1], Benjamin A. Frandsen[5], Bing Shen[1,*], Yusheng Hou[1,#],and Meng Wang[1,†]

[1]*Center for Neutron Science and Technology, Guangdong Provincial Key Laboratory of Magnetoelectric Physics and Devices, School of Physics, Sun Yat-Sen University, Guangzhou, Guangdong 510275, China*

[2] *School of Physics, Guangdong Province Key Laboratory of Display Material and Technology, Sun Yat-Sen University, Guangzhou, Guangdong 510275, China*

[3] *Key Laboratory of Bioinorganic and Synthetic Chemistry of Ministry of Education, School of Chemistry, Sun Yat-Sen University, Guangzhou, Guangdong 510275, China*

[4]*Beijing Synchrotron Radiation Facility, Institute of High Energy Physics, Chinese Academy of Sciences, Beijing 100049, China*

[5]*Department of Physics and Astronomy, Brigham Young University, Provo, Utah 84602, USA*

*# houysh@mail.sysu.edu.cn*

*\* shenbing@mail.sysu.edu.cn*

*† [wangmeng5@mail.sysu.edu.cn](mailto:wangmeng5@mail.sysu.edu.cn)*

## Properties of the single crystals of NiPSe$_3$ at atmospheric pressure

An X-ray diffraction (XRD) pattern of a single crystal of NiPSe$_3$ was measured at ambient pressure, as shown in Fig. S1. The XRD pattern shows (00L) diffraction peaks. The obtained lattice parameter $c$ is 6.8663(3) Å, which is consistent with the previous reports, confirming the high quality of the sample.

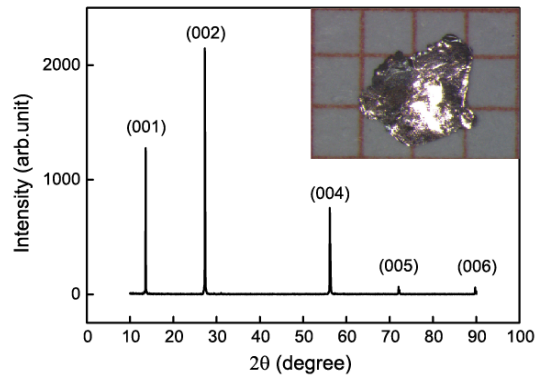

**Figure S1** XRD pattern of a single crystal of NiPSe$_3$ with the corresponding Miller indices (00$L$) at atmospheric pressure. The wavelength is $\lambda$=1.54 Å. The inset shows an image of a typical single crystal.

## Structural transitions of NiPSe$_3$ under pressure

The *in situ* high pressure XRD patterns of both the LP and HP-I phases can be well indexed by the monoclinic *C*2/*m* space group. During the isomorphic structural transition from LP to HP-I, there is an obvious sliding of the honeycomb layers relative to each other. The transition occurring at about 15.0 GPa is a non-isomorphic structural transition. At this transition, there is a large decrease of the interlayer spacing. The XRD patterns of the HP-II phase can be well indexed by the trigonal *P-31m* space group. The related structural parameters are listed in Table S1.

**Table S1** Refined lattice parameters, atomic coordinates, and Wyckoff positions (WP) of NiPSe$_3$ at 2.2, 6.3, and 25.7 GPa.

| \multicolumn{6}{c}{The LP phase at 2.2 GPa, space group: *C2/m*} |
|---|---|---|---|---|---|

| \multicolumn{6}{c}{$a = 6.052(1)$, $b = 10.440(1)$, and $c = 6.705(1)$ Å, $\alpha = 90°$, $\beta = 108.45(1)°$, $\gamma = 90°$, $R_{wp} = 2.04\%$, $R_p = 3.42\%$} |

| atom | x | y | z | Occ. | WP |
|---|---|---|---|---|---|
| Ni | 0 | 0.331(1) | 0 | 1 | 4g |
| P | 0.102(7) | 0 | 0.161(8) | 1 | 4i |
| Se 1 | 0.759(2) | 0 | 0.261(3) | 1 | 4i |
| Se 2 | 0.251(5) | 0.173(6) | 0.247(2) | 1 | 8j |

The HP-I phase at 6.3 GPa, space group: *C2/m*

$a = 5.898(1)$, $b = 10.348(1)$, and $c = 6.303(1)$ Å, $\alpha = 90°$, $\beta = 88.38(1)°$, $\gamma = 90°$, $R_{wp} = 5.12\%$, $R_p = 4.71\%$

| atom | x | y | z | Occ | WP |
|---|---|---|---|---|---|
| Ni | 0 | 0.3038(5) | 0 | 1 | 4g |
| P | 0.04(1) | 0 | 0.18(1) | 1 | 4i |
| Se 1 | 0.166(3) | 0.181(2) | 0.271(2) | 1 | 4i |
| Se 2 | 0.632(3) | 0 | 0.309(3) | 1 | 8j |

The HP-II phase at 25.7 GPa, space group: *P-31m*

$a = b = 5.873(2)$ and $c = 4.274(4)$ Å, $\alpha = 90°$, $\beta = 90°$, $\gamma = 120°$, $R_{wp} = 5.12\%$, $R_p = 4.71\%$

| atom | x | y | z | Occ | WP |
|---|---|---|---|---|---|
| Ni | 1/3 | 2/3 | 0 | 1 | 2c |
| P | 0 | 0 | 0 | 1 | 2e |
| Se | 0.327(3) | 0 | -0.330(2) | 1 | 6k |

**High-pressure resistance and Ginzburg-Landau fitting of the $\mu_0 H_{c2}$.**

The electrical resistance as a function of pressure and temperature for single crystals of NiPSe$_3$ is shown in Figs. S2a and S2b, respectively. The resistance decreases dramatically at the first structural transition and further decreases at ~8.0 GPa. The simple Ginzburg-Landau formula $\mu_0 H_{c2}(T) = \mu_0 H_{c2}(0)\left[1 - \left(\frac{T}{T_c}\right)^2\right]$ is adopted to fit the upper critical field of the superconductivity. The fitted values of $\mu_0 H_{c2}$ are 1.26 T at 8.0 GPa and 2.80 T at 34.0 GPa, lower than the Pauli limit.

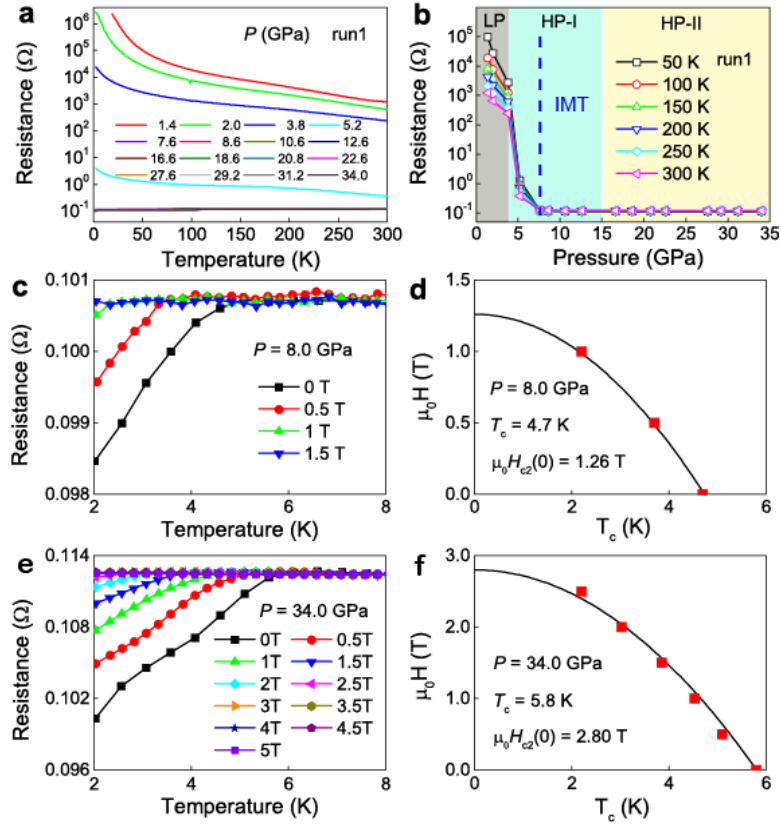

**Figure S2 High-pressure resistance measurements. a** Temperature dependence of the resistance under various pressures up to 34.0 GPa. The resistance is shown on a logarithmic scale. **b** Pressure dependence of the resistance at various temperatures up to 300 K shown on a logarithmic scale. The grey, blue and yellow backgrounds indicate the LP, HP-I, and HP-II phases. The vertical dashed line indicates the IMT at ~ 8.0 GPa. **c** and **e** Magnetic field dependence of the SC transition at 8.0 and 34.0 GPa. **d** and **f** Ginzburg-Landau formula fits to the experimentally determined $T_c$ as a function of magnetic field at 8.0 and 34.0 GPa.

## The AFM transition in NiPSe$_3$ under pressure

The suppressed $T_N$ of NiPSe$_3$ under pressure is due to the variation of the magnetic exchange couplings. At low pressure, NiPSe$_3$ is a Mott insulator with large resistance and its magnetic moment is around $2\mu_B/\text{Ni}^{2+}$. Because the AFM transition has an influence on the resistance, $T_N$ could be determined from the kink in the first derivative of the resistance. There is a dramatic decrease in the resistance at the first structural transition. After the structural transition, $T_N$ could be identified from the distinctive change in the resistance. Figure S3 displays selected temperature dependences of the resistance under various pressures from 2.0 to 7.6 GPa. The resistance curves are collected from two different measurements, including run 1 and run 2. The kinks indicate the values of $T_N$ that are marked in these plots. $T_N$ decreases from 203.1 K at 2.0 GPa to 139.2 K at 7.9 GPa. With further increasing pressure, the signal of the AFM transition becomes weaker and cannot be identified from the resistance. As shown in Fig. S4, the five nearest-neighbor exchange couplings are considered in NiPSe$_3$. Combining the DFT calculations and Monte Carlo simulations, the values of $T_N$ are determined accordingly. The results are listed in Table S2.

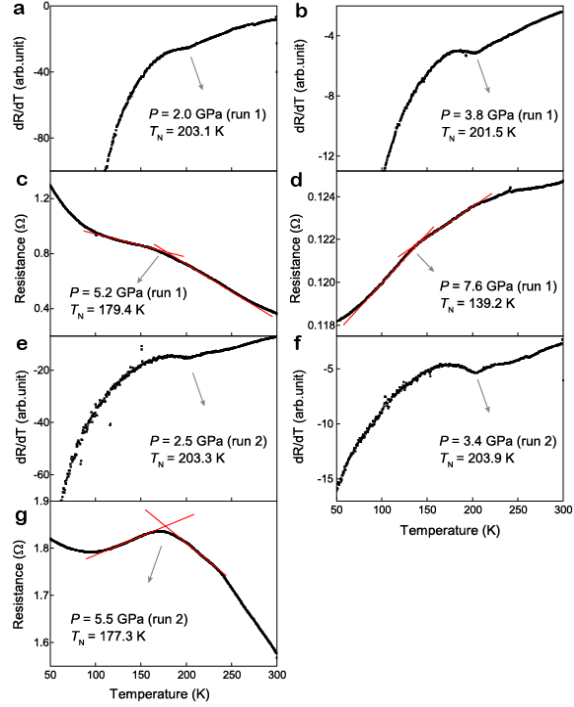

**Figure S3** Temperature dependence of the resistance under pressures of **a** 2.0, **b** 3.8, **c** 5.2, **d** 7.6, **e** 2.5, **f** 3.4, and **g** 5.5 GPa. The resistance curves are selected from different measurements (run 1 to 4). $T_N$ is determined by the inflection points marked in the plots.

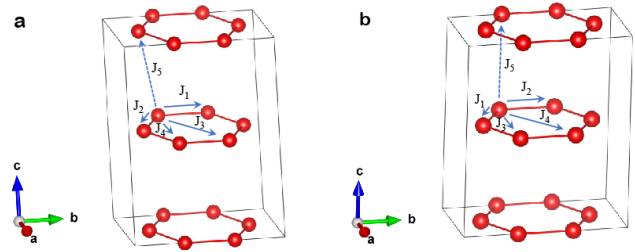

**Figure S4** Schematics of the Ni ions in the **a** LP phase and **b** HP-I phase.

**Table S2** DFT calculated magnetic exchange couplings $J$ in units of meV, electronic band gap, $T_N$, magnetic moment (M) and magnetic anisotropy energy (MAE) of NiPSe$_3$ under pressure.

| Pressure(GPa) | $J_1$ | $J_2$ | $J_3$ | $J_4$ | $J_5$ | Gap(eV) | $T_N$(K) | M($\mu_B$/Ni$^{2+}$) | MAE(meV/Ni$^{2+}$) |
|---|---|---|---|---|---|---|---|---|---|
| 0 | 7.90 | -2.89 | 1.47 | -1.03 | -7.41 | 0.994 | 189.87 | 1.264 | 0.991 |
| 10.4 | -8.66 | 53.9 | 1.71 | 2.64 | 3.72 | 0.004 | 142.40 | 0.866 | -0.074 |
| 12.3 | -7.49 | 4.42 | -2.52 | 2.97 | 3.35 | -0.168 | 122.76 | 0.772 | -0.091 |

As listed in Table S2, the interlayer magnetic exchange coupling $J_5$ changes from AFM to FM with pressure increasing. Thus, we establish a 1×1×2 out-of-plane supercell to deliberate the magnetic ground state. As shown in Fig.S5, the zigzag-out magnetic structure is more stable than the zigzag-in structure when the pressure on NiPSe$_3$ is less than 6 GPa. However, zigzag-out has higher energy than zigzag-in when the pressure is larger than 6 GPa. When the pressure is above 15 GPa, NiPSe$_3$ favors a non-magnetic phase.

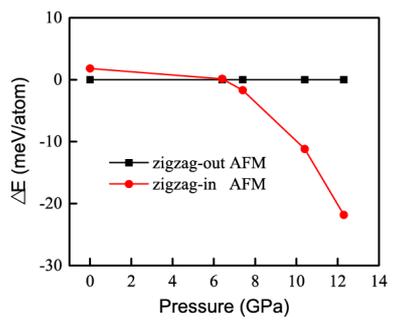

**Figure S5** Pressure dependence of the energy difference between the zigzag-out and zigzag-in AFM orders of NiPSe$_3$. Here, the energies of the zigzag-out AFM order are taken as reference.